%% file: cernrep.tex
\documentclass{cernrep} 
\usepackage{texnames}
\usepackage[T1]{fontenc}
 \def\bc{\begin{center}}          \def\ec{\end{center}}
\usepackage{subfig}
\usepackage{graphicx}
\usepackage{rotating}
\usepackage{authblk}

\title{Particle physics applications of the AWAKE acceleration scheme}

\vspace{2cm}

\author[1]{A.~Caldwell}
\author[2]{J.~Chappell}
\author[3]{P.~Crivelli}
\author[3]{E.~Depero}
\author[4]{J.~Gall}
\author[5]{S.~Gninenko}
\author[4]{E.~Gschwendtner}
\author[2]{A.~Hartin}
\author[2]{F.~Keeble}
\author[4]{J.~Osborne}
\author[4]{A.~Pardons}
\author[4]{A.~Petrenko}
\author[2]{A.~Scaachi}
\author[2]{M.~Wing\footnote{Corresponding author: m.wing@ucl.ac.uk}}
\affil[1]{Max Planck Institute for Physics, Munich, Germany}
\affil[2]{University College London, London, UK}
\affil[3]{ETH Z\"{u}rich, Switzerland}
\affil[4]{CERN, Geneva, Switzerland}
\affil[5]{INR Moscow, Russia}

\begin{document}

\maketitle

\begin{abstract}{
The AWAKE experiment had a very successful Run 1 (2016--8), demonstrating proton-driven plasma wakefield 
acceleration for the first time, through the observation of the modulation of a long proton bunch into micro-bunches and the 
acceleration of electrons up to 2\,GeV in 10\,m of plasma.  The aims of AWAKE Run 2 (2021--4) are to have high-charge 
bunches of electrons accelerated to high energy, about 10\,GeV, maintaining beam quality through the plasma and 
showing that the process is scalable.  The AWAKE scheme is therefore a promising method to accelerate electrons to 
high energy over short distances and so develop a useable technology for particle physics experiments.  Using proton 
bunches from the SPS, the acceleration of electron bunches up to about 50\,GeV should be possible.  Using the LHC 
proton bunches to drive wakefields could lead to multi-TeV electron bunches, e.g.\ with 3\,TeV acceleration achieved 
in 4\,km of plasma.  This document outlines some of the applications of the AWAKE scheme to particle physics and 
shows that the AWAKE technology could lead to unique facilities and experiments that would otherwise not be possible.  
In particular, experiments are proposed to search for dark photons, measure strong field QED and investigate new physics 
in electron--proton collisions.  The community is also invited to consider applications for electron beams up to the TeV scale.
}
\end{abstract}

\vspace{2cm}
\centerline{\bf Input to the European Particle Physics Strategy Update}
\vspace{3cm}

\centerline{December 2018}
\vspace{2cm}

\pagebreak[4]
\


\input{introduction}


\end{document}

%% file: introduction.tex
\section{Introduction}

The AWAKE experiment shows great promise to be able to provide electron beams at high energy ranging up to the TeV scale.  The 
experiment and its plans are described in another submission to this process~\cite{AWAKE-sub} and assuming AWAKE continues to successfully 
demonstrate the technique, particle physics applications of the AWAKE scheme can already be considered.  The advantage of 
using proton-driven plasma wakefield acceleration as used by AWAKE is the promise to be able to accelerate electrons to high 
energy in one acceleration stage with relatively high gradients of $\mathcal{O}$(GV/m).  This means that using proton bunches from 
the SPS, the acceleration of electron bunches up to about 50\,GeV should be possible.  Using the LHC proton bunches to drive 
wakefields could lead to multi-TeV electron bunches, e.g.\ with 3\,TeV acceleration achieved in 4\,km of plasma~\cite{pp:18:103101}.

Currently, high energy electrons ($> 50$\,GeV) are only possible as part of the secondary SPS beam and at low rate.  Therefore, 
using the AWAKE scheme would provide the highest energy, high-charge electron bunches in the world.  The 
acceleration of electrons to high energies opens up the possibility of new particle physics experiments~\cite{wing:royalsoc}: the 
search for dark photons (see Section~\ref{sec:dark-photon}); measurement of quantum electrodynamics (QED) in strong fields (see Section~\ref{sec:sfqed}); and high energy electron--proton collisions (see Section~\ref{sec:ep}).   Further experimental ideas may 
emerge and the community is invited to consider possible uses of high-energy electron beams.  In parallel, the AWAKE technology 
will be further developed which is critical for any application to particle physics.   The integration and technical 
issues related to realising these experiments within the CERN site are part of a separate submission~\cite{AWAKE-technical}.

\section{AWAKE experiment}
\label{sec:awake}

The AWAKE experiment had a tremendously successful Run 1 (2016--8), demonstrating proton-driven plasma wakefield 
acceleration for the first time, through the observation of the modulation of a long proton bunch into 
micro-bunches~\cite{awake1,awake2} and the acceleration of electrons up to 2\,GeV in 10\,m of 
plasma~\cite{awake3}.  

The AWAKE Run 1 programme finished in November 2018 along with all experiments that rely on the CERN proton accelerators.  
The SPS will start up again in 2021 and run for four years until 2024 and an ambitious AWAKE Run 2 programme is being developed 
for this period.  The aims of AWAKE Run 2 are to have high-charge bunches of electrons accelerated to high energy, about 
10\,GeV, maintaining beam quality through the plasma and showing that the process is scalable.  This will require development 
of the initial electron source, beam and plasma diagnostics as well as development of the plasma technology which can fulfil 
these ambitious goals.  The final goal by the end of AWAKE Run 2 is to be in a position to use the AWAKE scheme for particle 
physics experiments.  

\section{Possible particle physics experiments}
\label{sec:expts}

A high-energy electron beam with high-charge bunches from tens of GeV up to TeV energies has many potential 
applications.  Three are outlined in detail in the following sub-sections, but other possibilities are briefly discussed 
here.  A condition of any application is that the particle physics goals must be new, interesting and do something 
not done elsewhere.  An oft mooted application of plasma wakefield acceleration is the development of a high energy, 
high luminosity linear $e^+e^-$ collider.  However, such a collider is a challenge for all accelerator technology and 
so to have this as the first application of plasma wakefield acceleration is ambitious.  Hence, the approach taken here 
is to consider experiments, such as fixed-target experiments and an electron--proton collider, which have less 
stringent requirements on the quality of the beam.  A natural progression would be to build such accelerators before 
attempting to develop a high energy, high luminosity linear $e^+e^-$ collider.  In this way, the accelerator technology 
can be developed whilst still carrying out cutting-edge particle physics experiments.

A high-energy electron beam could be used as a test-beam facility for either detector or accelerator studies.  There 
are not many such facilities world-wide and they are often over-subscribed.  The characteristics of the electron 
beam which would make it rather distinct are: the high energy which can be varied; a pure electron beam with low 
hadronic backgrounds; a high bunch charge; and longitudinally short bunches.  These properties may not be ideal 
or will be challenging for detector studies which usually rely on single particles.  However, as an accelerator test 
facility the bunched structure and flexibility in energy will be useful properties.

The AWAKE scheme could be used to accelerate bunches of muons to high energies with small losses through decay 
of the muons.  In the recently proposed scheme for the front end of a muon collider~\cite{muon-collider}, bunches of 
$6 \times 10^9$\,$\mu^+$ and $\mu^-$ are produced with $\sigma_z =100$\,$\mu$m and $\sigma_r =40$\,$\mu$m.  
Initial estimates indicate that such bunches of muons could be effectively accelerated given electric fields of 2\,GV/m, 
which can be achieved with LHC proton bunches.

The future circular collider (FCC) would be an excellent driver of plasma wakefields given the very high proton energy 
and the small bunch emittance~\cite{fcc}.  Introducing long plasma cells in the straight sections of the FCC could lead to  
the production of multi-TeV electron bunches, further greatly extending the physics capability of the FCC.  It may also 
be possible to accelerate electron bunches to 50\,GeV or more in the straight sections without significant loss of protons, 
thus allowing for a high luminosity $ep$ or $e^+e^-$ programme at moderate additional cost.

There are most certainly other possible applications of proton-driven plasma wakefield acceleration, encapsulated in the 
AWAKE scheme, that could be proposed and investigated.  This document represents some compelling ideas, with the 
following sections discussing the most developed ideas.

\subsection{Dark photon experiment}
\label{sec:dark-photon}

Dark photons~\cite{JETP:56:502,pl:b136:279,pl:b166:196} are postulated particles which could provide the link to a dark or hidden sector 
of particles.  This hidden sector could explain a number of issues in particle physics, not least of which is that they are candidates 
for dark matter which is expected to make up about 80\% of known matter in the Universe.  Dark photons are expected to have low masses 
(sub-GeV)~\cite{pl:b513:119,pr:d79:015014} and couple only weakly to the Standard Model particles and so would have not been seen in 
previous experiments.  
The dark photon, labelled $A^\prime$, is a light vector boson which results from a spontaneously broken new gauge symmetry 
and kinetically mixes with the photon and couples to the electromagnetic current with strength $\epsilon \ll 1$.  Recently, experimental 
and theoretical interest in the hidden sector has increased and is discussed in recent reviews on the 
subject~\cite{arnps:60:405,us-cosmic-visions}.  

\begin{figure}[h]
\centering
\includegraphics[scale=0.7]{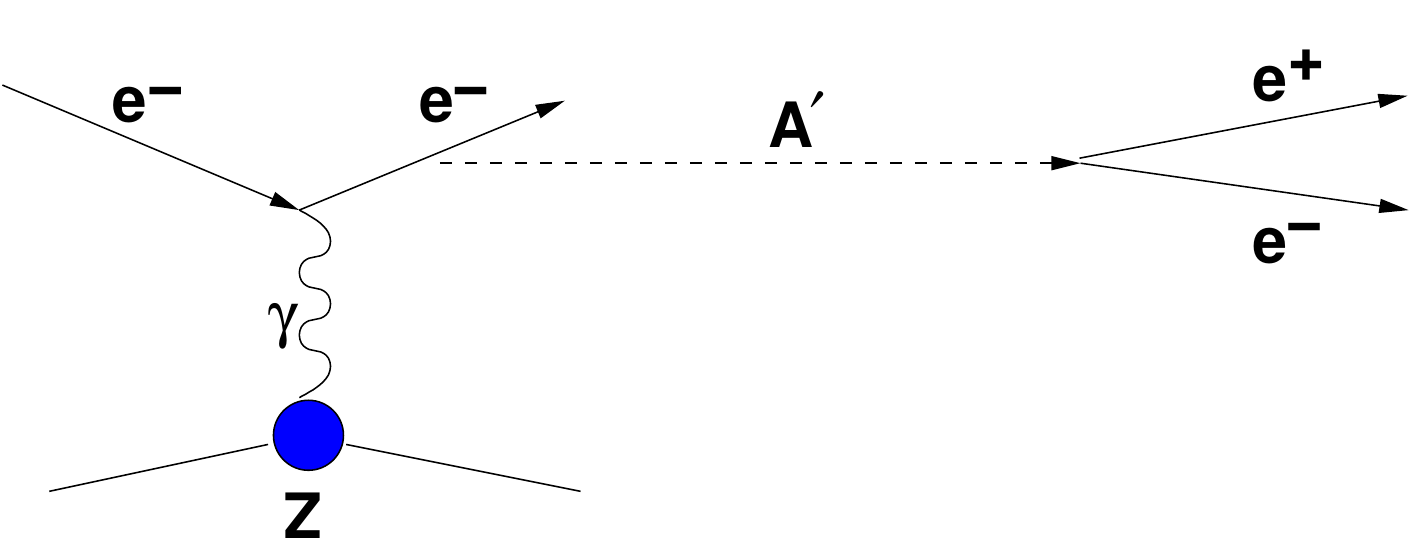}
\caption{A representation of the production of a dark photon, $A^\prime$, in a fixed-target experiment with an electron beam.  
The dark photon subsequently decays to an $e^+e^-$ pair.} 
\label{fig:dark_photon_process}
\end{figure}

A common approach to search for dark photons is through the interaction of an electron with a target in which the dark photon is 
produced and subsequently decays.  This process is shown in Fig.~\ref{fig:dark_photon_process} in which the dark photon decays to an 
$e^+e^-$ pair.  The NA64 experiment is already searching for dark photons using high-energy electrons on a 
target~\cite{pr:d89:075008,arxiv:1312.3309,prl:118:011802}, initially measuring the dark photon decaying to dark matter particles 
(``invisible mode'')
and so leaving a signature of missing energy in the detector.  Although high-energy electrons of 100\,GeV are used, a limitation of 
the experiment is that the rate of electrons is about $10^6$\,electrons per second as they are produced in secondary interactions 
of the SPS proton beam.

\begin{figure}[ht]
\centering
\includegraphics[scale=0.7]{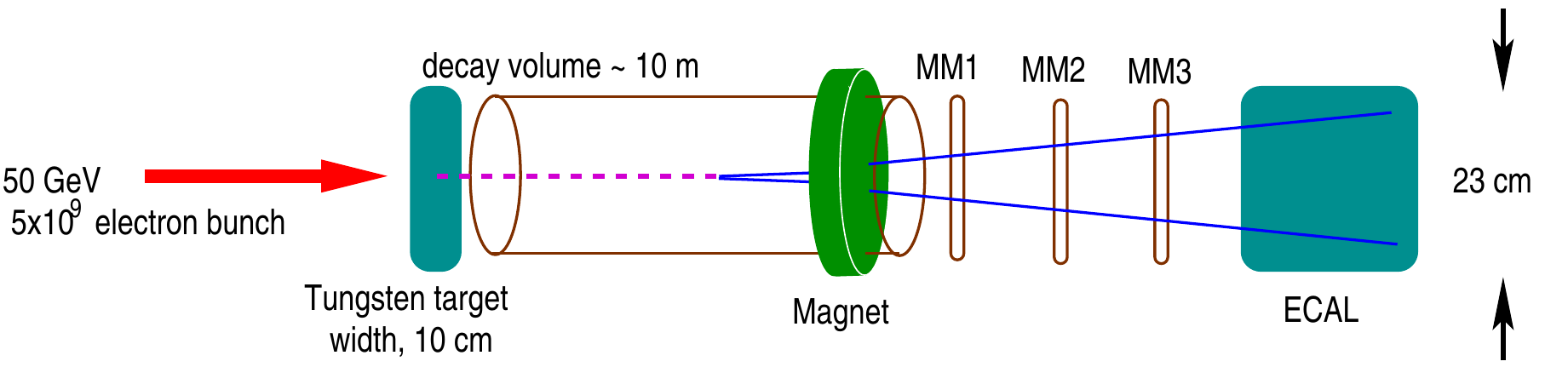}
\caption{A sketch of the experimental setup for a bunch of $5 \times 10^9$ electrons each of 50\,GeV produced via the AWAKE 
scheme impinging on a tungsten target of depth 10\,cm.  The target is followed by a decay volume and a dipole magnet to 
separate the electrons and positrons which are then tracked through three tracker planes (MM1, MM2 and MM3), followed by 
an electromagnetic calorimeter (ECAL).} 
\label{fig:AWAKE-NA64}
\end{figure}

Based on a {\sc Geant4}~\cite{geant} simulation of the NA64 experiment, the setup has been modified to cater for the use of electron bunches 
rather than single electrons.  Default physics processes, including dark photon production and decay, are implemented in the {\sc Geant4} 
simulation.  A schematic of the setup is shown in Fig.~\ref{fig:AWAKE-NA64} where the main components are 
shown.  A 10\,cm long tungsten target is followed by a 10\,m long volume in which the dark photon can decay.  The decay 
products, the $e^+e^-$ pair, are then separated via a dipole magnet and detected in micromegas tracker planes (MM1, MM2 and 
MM3), followed by a tungsten--plastic shashlik electromagnetic calorimeter (ECAL).  The analysis is based on well-separated 
hits ($> 1$\,mm) in the three tracker planes, as well as low background rates from the optimised target thickness.  
Therefore a dark photon decay is determined via detection of the decay products, reconstruction of a displaced vertex and reconstruction 
of the $A^\prime$ invariant mass. 
The sensitivity to dark photon production is evaluated at 90\% confidence level in the $\epsilon-m_{A^\prime}$ plane, assuming 
a background-free case and an overall signal reconstruction efficiency of $\sim$50\%.

\begin{figure}[ht]
\centering
\includegraphics[width=12cm,trim={5.cm 0cm 5.cm 1.5cm},clip]{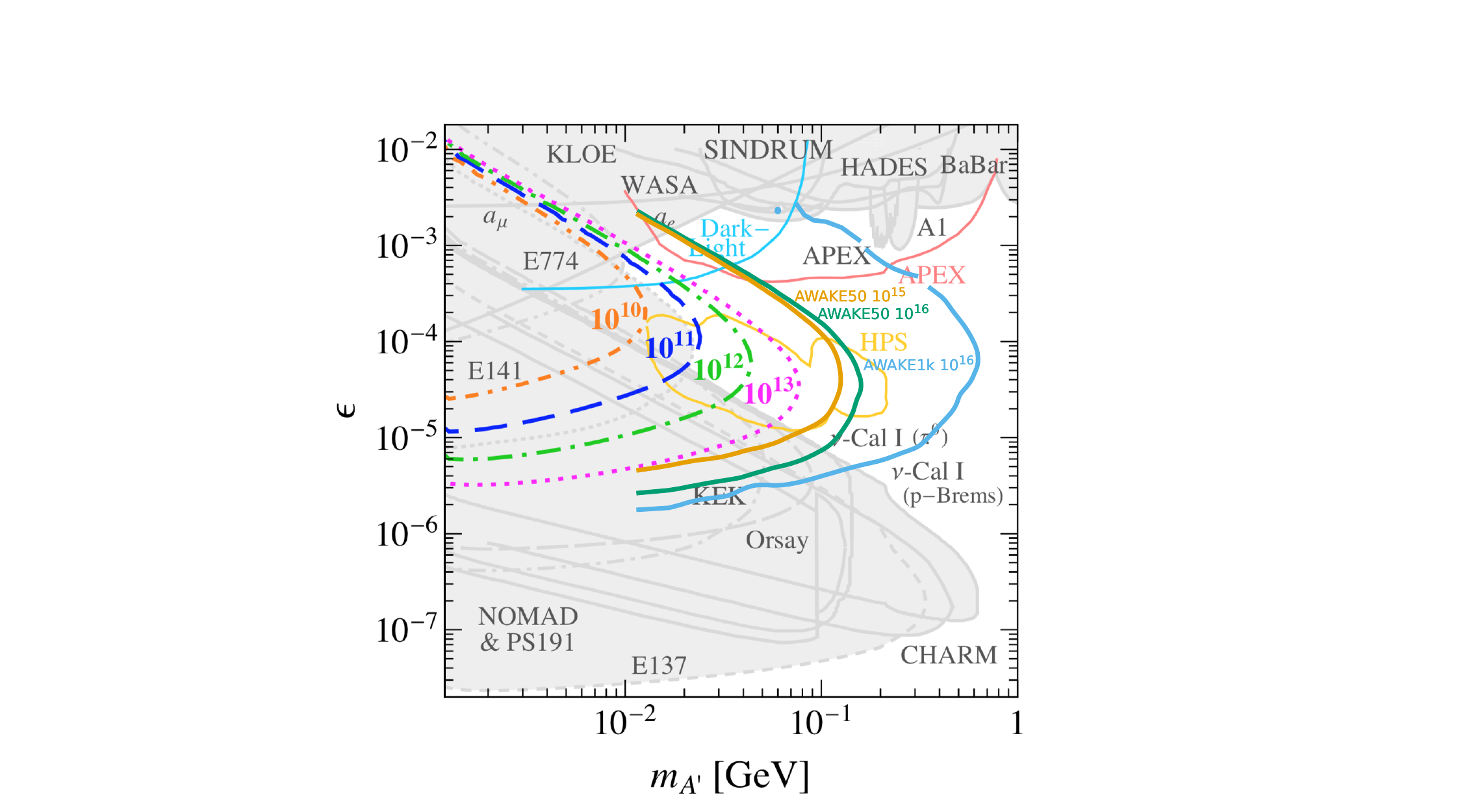}
\caption{Limits on dark photon production decaying to an $e^+e^-$ pair in terms of the mixing strength, $\epsilon$ and dark 
photon mass, $m_{A^\prime}$, from previous measurements (light grey shading).  The expected sensitivity for the NA64 
experiment is shown for a range of electrons on target, $10^{10} - 10^{13}$.  Expectations from other potential experiments 
are shown as coloured lines.  Expected limits are also shown for $10^{15}$ (orange line) or $10^{16}$ (green line) electrons of 
50\,GeV (``AWAKE50'') on target and $10^{16}$ (blue line) electrons of 1\,TeV (``AWAKE1k'') on target 
provided to an NA64-like experiment by a future AWAKE accelerator scheme; these are the the results 
of work performed here.} 
\label{fig:dark_photon_limit} 
\end{figure}

The NA64 experiment is, however, already making significant progress investigating new regions of phase space for dark photons and 
as shown in Fig.~\ref{fig:dark_photon_limit} will cover much new ground in the $\epsilon-m_{A^\prime}$ plane.  Given the 
limitations of the number of electrons on target, the AWAKE acceleration scheme could make a real impact as the number 
of electrons is expected to be several orders of magnitude higher.  Assuming a bunch of $5 \times 10^9$ electrons and a 
running period of 3\,months gives $10^{16}$ electrons on target and this is shown in Fig.~\ref{fig:dark_photon_limit}; to visualise 
the effect of the number of electrons on target, the expectation for $10^{15}$ electrons is also shown.  Our results 
in the figure clearly show that we will be able to probe a new region, in particular extending to higher masses in the region 
of $10^{-3} < \epsilon < 10^{-5}$.  Also shown in the figure are results using bunches of electrons, each of energy 1\,TeV, again 
with $10^{16}$ electrons on target.  Such a search could be part of a future collider programme, e.g.\ a very high energy 
$ep$ collider (discussed in Section~\ref{sec:ep}), in which active use of the beam dump is made.  The higher energy electron beam 
extends the sensitivity significantly to higher mass dark photons, covering a region unexplored by current or planned experiments, 
between the regions covered by current colliders and previous high intensity beam-dump experiments.

Further studies are ongoing and a higher number of electrons on target should be possible depending on the SPS injection 
scheme as well as the success of AWAKE in accelerating bunches of electrons.  An optimised detector configuration will be 
investigated, as will other decay channels, such as $A^\prime \to \mu^+ \mu^-$ or $A^\prime \to \pi^+ \pi^-$ as well as the 
invisible modes, and effects of the beam energy.  Such an experiment could be realised during and after LS3 in extensions 
of the current AWAKE area; technical studies of this possibility and infrastructure requirements are discussed 
elsewhere~\cite{AWAKE-technical}.

\subsection{Strong-field quantum electrodynamics}
\label{sec:sfqed}

The theory of electromagnetic interactions, QED, has been studied and tested in numerous 
reactions, over a wide kinematic range and often to tremendous precision.  The collision of a high-energy electron 
bunch with a high-power laser pulse creates a situation where QED is poorly tested, namely in the strong-field regime.  In 
the regime around the Schwinger critical field, $\sim 1.3 \times 10^{18}$\,V/m, QED becomes non-linear and these values 
have so far never been achieved in controlled experiments in the laboratory.  Investigation of this regime could lead to a 
better understanding of where strong fields occur naturally such as on the surface of neutron stars, at a black hole's event 
horizon or in atomic physics.

\begin{figure}[!h]
\centering\includegraphics[width=8cm,trim={7cm 12cm 7cm 0cm},clip]{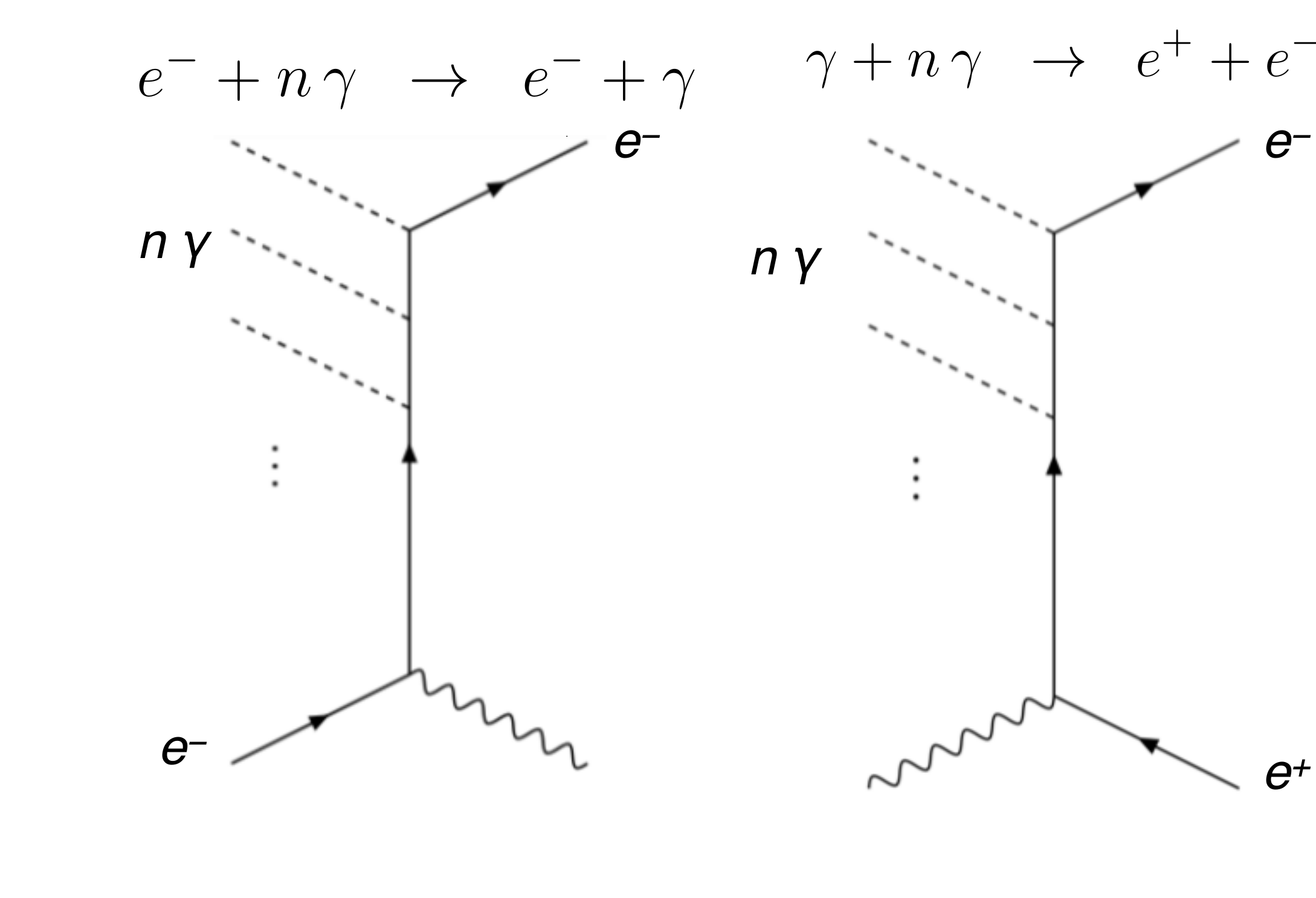}
\caption{A Feynman diagram representation of (left) Compton scattering of an electron and (right) production of an 
$e^+e^-$ pair in the field of a high-power laser in which absorption of multiple photons has taken place.}
\label{fig_qed}
\end{figure}

In the presence of strong fields, rather than the simple $2 \to 2$ particle scattering, e.g.\ $e^- + \gamma \to e^- + \gamma$,  
multi-particle absorption in the initial state is possible, e.g.\ $e^- + n\gamma \to e^- +\gamma$, where 
$n$ is an integer (see Fig.~\ref{fig_qed}).  Therefore an electron interacts with multiple photons in the laser pulse and a photon can also 
interact with multiple photons in the laser pulse to produce an $e^+e^-$ pair, also shown in Fig.~\ref{fig_qed}.  For more details 
on the processes and physics, see a recent review~\cite{hartin}.

\begin{figure}[ht]
\centering
\includegraphics[scale=0.33]{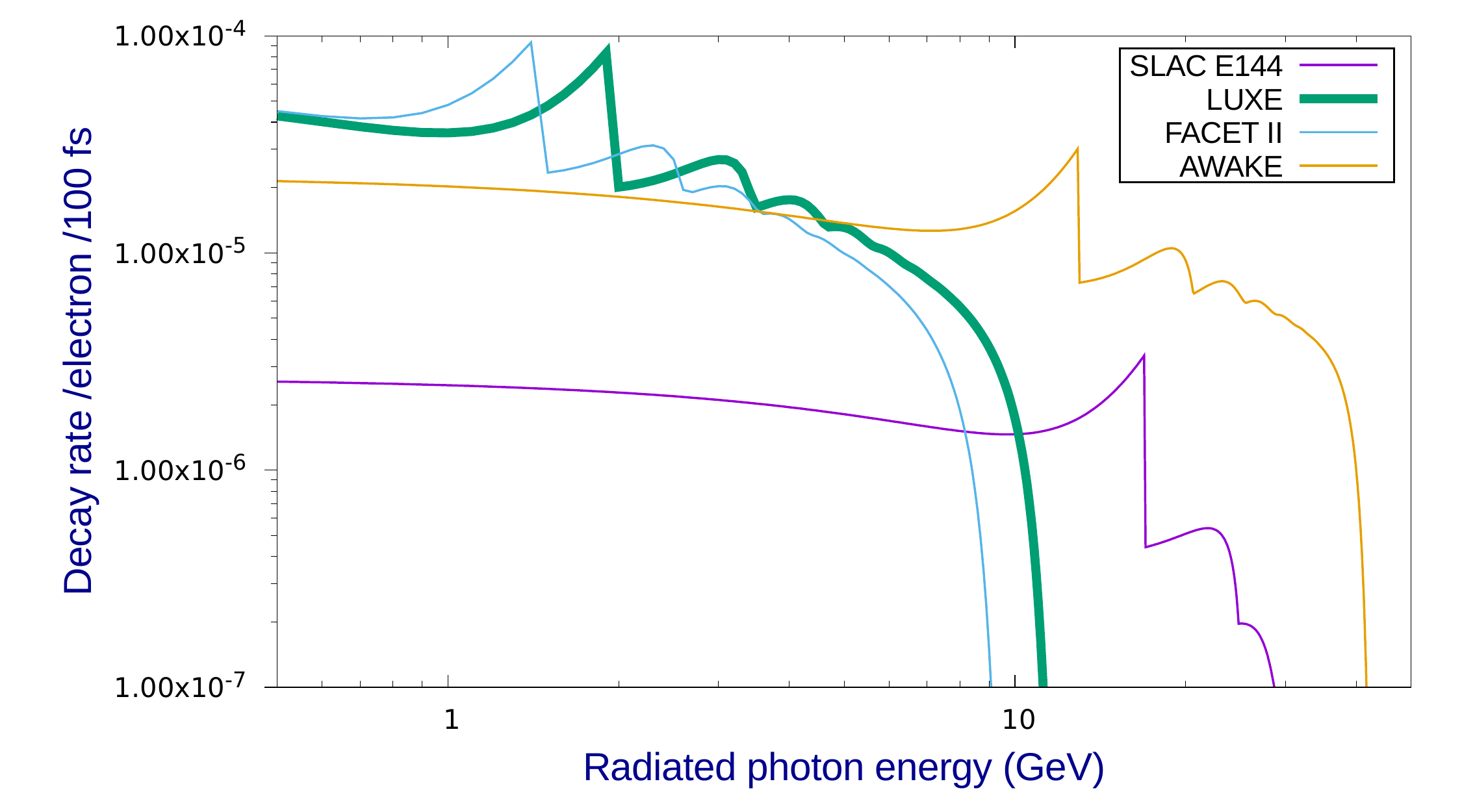}
\includegraphics[scale=0.33]{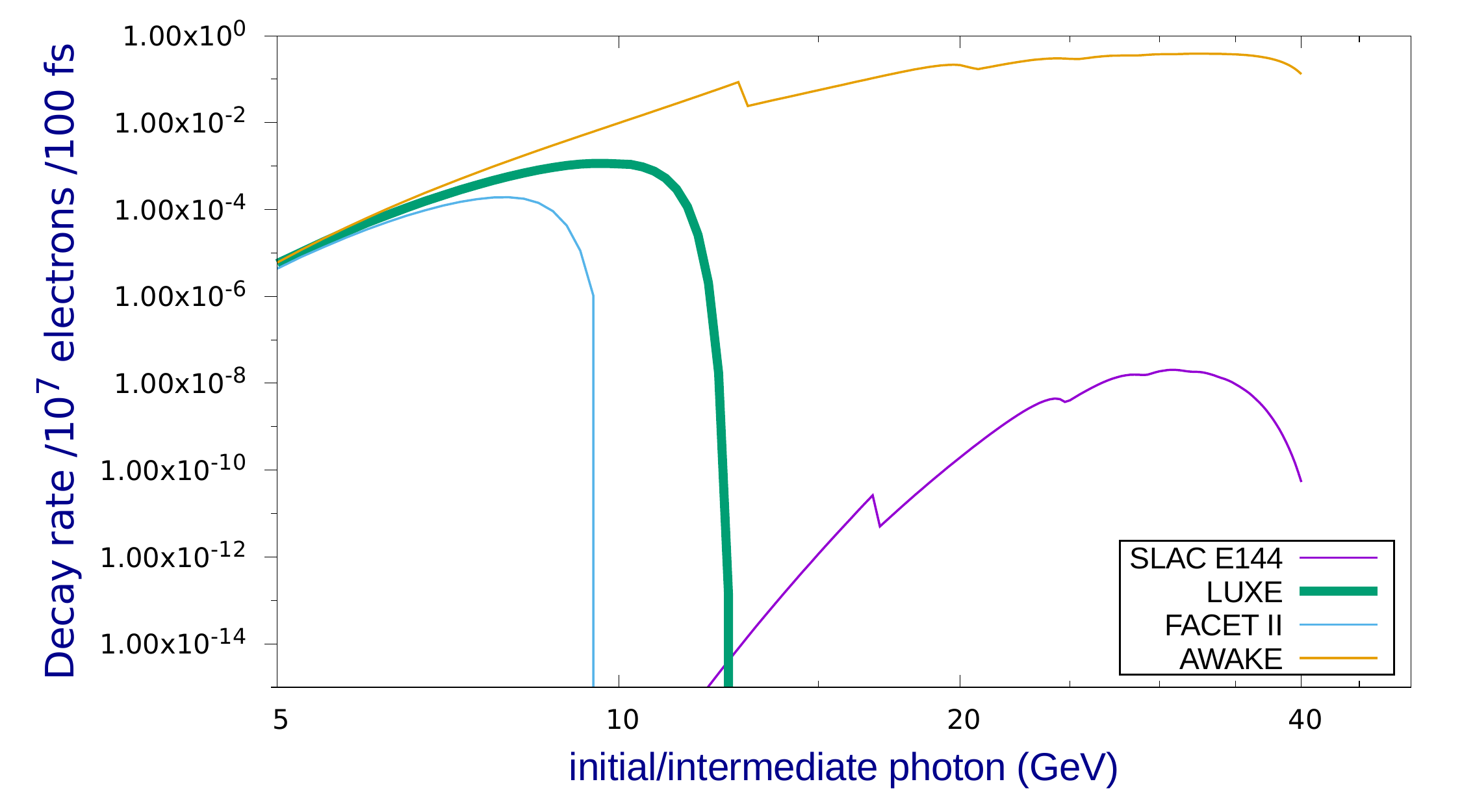}
\caption{(Left) energy of produced photons in reaction $e^- + n\gamma \to e^- +\gamma$ and (right) energy of photons 
in the pair production process, $\gamma + n\gamma \to e^- + e^+$, where $n$ is an integer.  Results are shown for the only 
experiment that has so far probed close to this strong-field regime (SLAC E144), two proposed experiments (LUXE and 
FACET II) and what would be possible with 50\,GeV electrons from AWAKE.} 
\label{fig:qed} 
\end{figure}

The E144 experiment~\cite{pr:d60:092004} at SLAC investigated electron--laser collisions in the 1990s using bunches of 
electrons, each of energy about 50 GeV, but due to the limitations of the laser, they did not reach the Schwinger critical 
field in the rest frame of the electrons.  With the advances in laser technology over the last 20\,years, these strong fields 
are now in reach~\cite{sqed-ws}.  However, the current highest-energy bunches of electrons of high charge are delivered 
by the European XFEL at 17.5\,GeV and the AWAKE scheme has the possibility to provide a higher-energy electron beam 
which would then be more sensitive to the $e^+e^-$ pair production process and probe a different kinematic regime.  This 
is shown in Fig.~\ref{fig:qed}, where the production rates for an AWAKE beam are much higher, particularly at high energies.  
AWAKE is compared to E144 and two potential experiments at FACET II and LUXE; typical parameters for the experiments 
are given in Table~\ref{table:sfqed}.

\begin{table}[!h]
\begin{center}
\caption{Laser and electron bunch parameters achieved for the E144 experiment.  Typical parameters are shown for planned 
experiments; they will use much shorter pulses leading to higher power than E144.}
\label{table:sfqed}
\begin{tabular}{l|cccc}
\hline
Parameter & E144 & LUXE & FACET II & AWAKE \\
\hline
Laser wavelength (nm) & 527/1053 & 527/1053 & 527/800/1053 & 527 \\
Laser energy (J) & 2 & 2 & 1 & 1 \\
Laser transverse size ($\mu$m$^2$) & 50 & 100 & 64 & 64 \\
Laser pulse length (ps) & 1.88 & 0.05 & 0.04 & 0.04 \\
Electron energy (GeV) & 46.6 & 17.5 & 15 & 50 \\
Electrons per bunch & $5 \times 10^9$ & $6 \times 10^9$ & $5 \times 10^9$ & $5 \times 10^9$ \\  \hline
\end{tabular}
\end{center}
\end{table}

With the higher rates possible for strong field pair production with AWAKE electrons, the Schwinger tunnelling regime will be 
reached in the rest frame. The signature of this regime will be an asymptote to a simple exponential dependence on the 
Schwinger critical field. This makes accessible the first experimental measurement of the Schwinger critical field 
strength~\cite{arXiv:1807.10670}. 

\subsection{High energy electron--proton collisions}
\label{sec:ep}

A natural avenue of study with a high-energy electron beam is the deep inelastic scattering of electrons off protons 
or ions in order to study the fundamental structure of matter~\cite{dis:book}.  The simplest experimental configuration 
is where a lepton beam impinges on a fixed target and many such experiments have been performed in the past.  So 
far only one lepton--hadron collider, HERA~\cite{hera}, has been built.  Potential physics that could be studied at a 
future deep inelastic scattering fixed-target experiment are to measure the structure of the proton at high momentum fraction of 
the struck parton in the proton, which could be valuable for the LHC, and to understand the spin structure of the nucleon, 
which is still poorly known~\cite{mpl:a24:1087}.  

A thorough survey of previous as well as planned experiments must be carried out to assess the potential of a deep 
inelastic scattering fixed-target experiment based on a high-energy electron beam of ${\mathcal O}(50)$\,GeV from AWAKE.  
The use of TeV electron beams would lead to fixed-target experiments at significantly higher energies ($\sqrt{s} = 75$\,GeV 
for a 3\,TeV electron beam) and comparable to the next generation nuclear physics collider, the EIC, which expects 
centre-of-mass energies between 20 and 140\,GeV.

A high-energy electron--proton/ion ($ep/eA$) facility could be the first application of plasma wakefield acceleration to particle 
colliders.  In such collisions, the electron generally emits a photon of virtuality $Q^2$ and strikes a parton carrying a fraction, 
$x$, of the proton's momentum.  The higher the $Q^2$, the smaller the probe and hence a more detailed structure can be 
seen; also low values of $x$ probe low momentum particles and hence the dynamic structure of quark and gluon radiation 
within the proton.  As such, $ep/eA$ collisions provide a detailed picture of the fundamental structure of matter and 
investigate the strong force of nature and its description embodied within quantum chromodynamics (QCD).  Some of the 
open issues to be investigated in $ep/eA$ collisions are: when does this rich structure of gluon and quark radiation stop or 
"saturate" as it surely must otherwise cross sections would become infinite; in general, the nature of high-energy hadronic 
cross sections; and is there further substructure or are partons fundamental point-like objects.

Initial collider designs~\cite{guoxing} considered generating electron bunches via the AWAKE scheme with electrons up to about 100\,GeV.  
This has been formulated into the PEPIC (Plasma Electron--Proton/Ion Collider) project in which the SPS protons are used 
to drive wakefields and accelerate electrons to about $50-70$\,GeV which then collide with LHC protons.  Given the high-luminosity 
LHC beam parameters and the rate of SPS bunches, the luminosity for PEPIC is currently estimated to be about 
$1.5 \times 10^{27}$\,cm$^{-2}$\,s$^{-1}$~\cite{AWAKE-technical}.  Therefore, PEPIC would have 
essentially the same energy reach as the LHeC project, but with luminosities several orders of magnitude lower.  Given integrated 
luminosities of about 10\,nb$^{-1}$ per year, it would focus on studies of the structure of matter and QCD in a new kinematic domain, 
in particular at low values of $x$, and total cross sections where the 
event rates are high.  This collider would be an option for CERN should the LHeC not be realised.  Investigation of increasing the 
luminosity through larger electron bunch population, smaller proton bunch size and higher SPS bunch frequency should also be pursued.

A very high energy electron--proton (VHEeP) collider has been proposed~\cite{epj:c76:463} in which LHC bunches are used 
to drive wakefields and accelerate electrons to 3\,TeV in under 4\,km, which then collide with the counter-propagating proton 
(or ion) bunch, creating electron--proton collisions at centre-of-mass energies, $\sqrt{s}$, of over 9\,TeV.  The energies of the 
electrons could be varied; the distance of 4\,km fits comfortably within the circumference of LHC ring, so although 
there may be an upper energy limit, lower energies should be achievable.  Such centre-of-mass energies represent a factor 
of 30 increase compared to HERA which allows an extension to low $x$ and to high $Q^2$ of a factor of 1000.  The luminosity 
is currently estimated to be around $10^{28} - 10^{29}$\,cm$^{-2}$\,s$^{-1}$ which would lead to an integrated luminosity of 
1\,pb$^{-1}$ per year.  Different schemes to improve this value are being considered such as squeezing the proton (and 
electron) bunches, multiple interaction points, etc..  However, even at these modest luminosities, such a high-energy 
electron--proton collider has a strong physics case.

\begin{figure}[!h]
\centering\includegraphics[width=7cm]{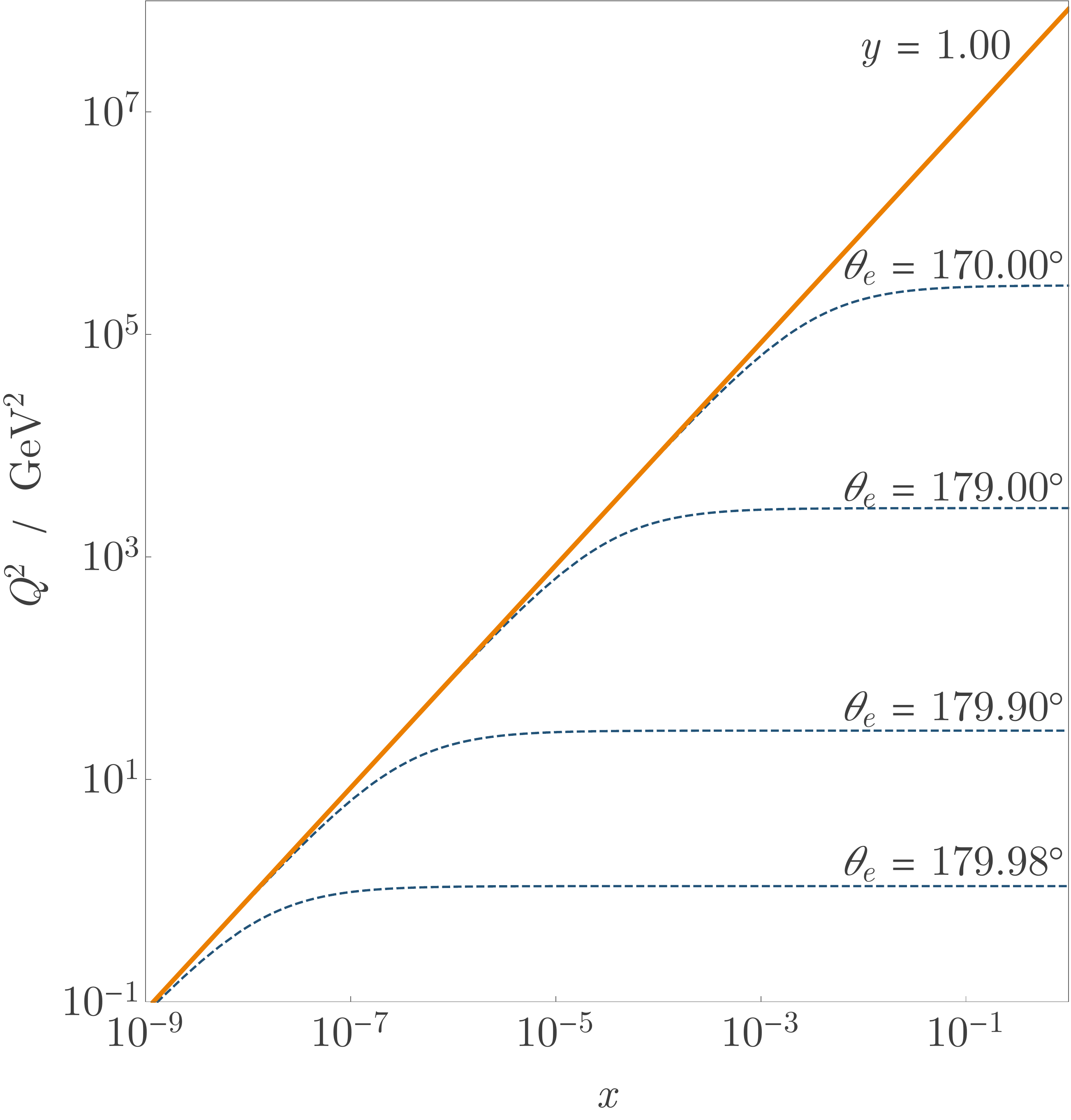}
\centering\includegraphics[width=7cm]{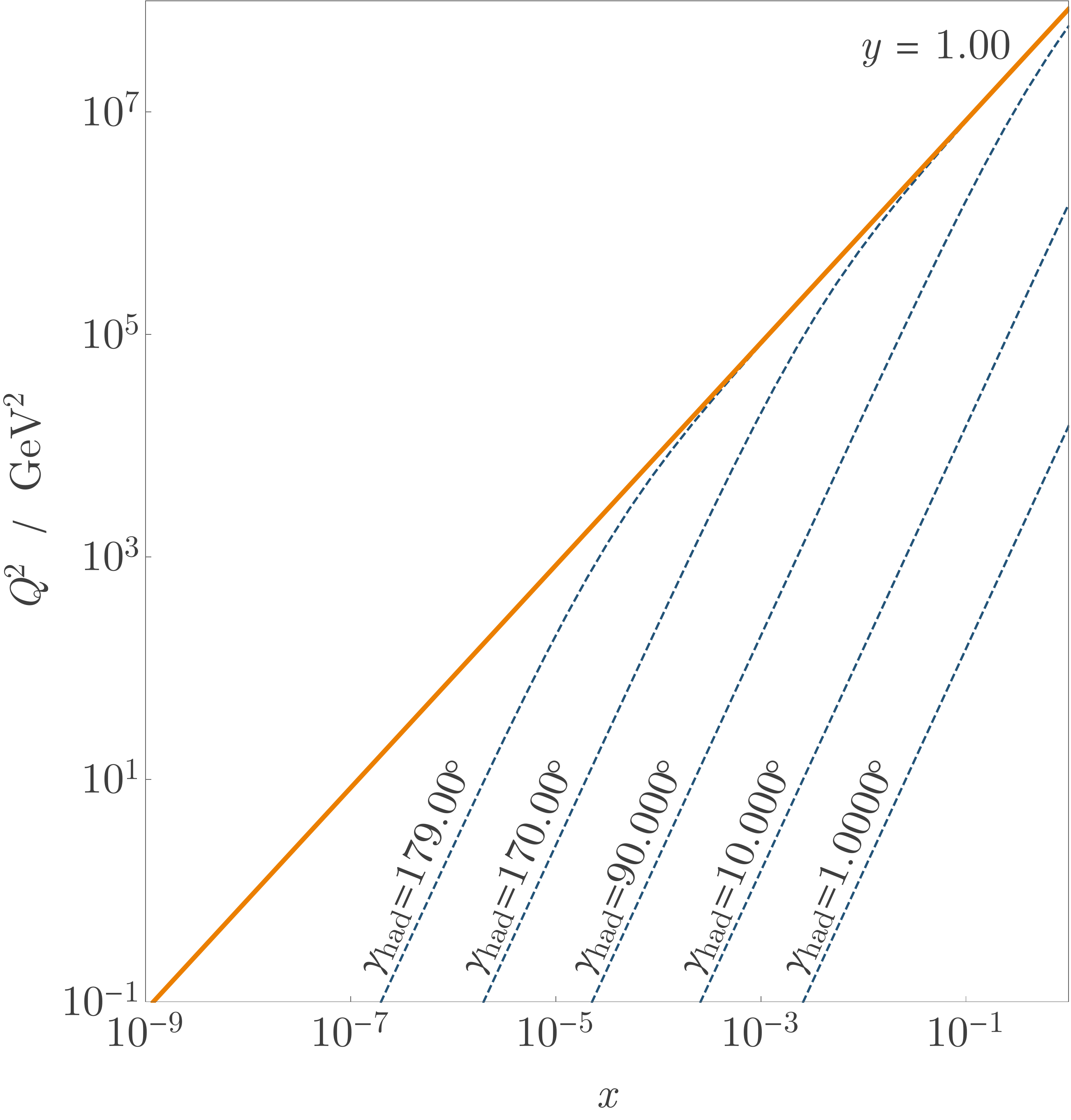}
\caption{The accessible $Q^2$ and $x$ coverage for VHEeP with $\sqrt{s} = 9$\,TeV with the kinematic limit of the 
inelasticity variable $y=1$ shown.  
Also shown are (left) lines indicating the electron scattering angle and (right) lines indicating the angle of the final-state hadronic 
system,  where $\theta_e =0, \gamma_{\rm had}=0$ indicates the direction of the proton beam.}
\label{fig:vheep_kinematics}
\end{figure}

The kinematic reach in $Q^2$ and $x$ for VHEeP is shown in Fig.~\ref{fig:vheep_kinematics}, with e.g.\ a minimum requirement 
of $Q^2 =1$\,GeV$^2$ corresponding to a minimum value of $x \sim 10^{-8}$.  At such values, even with integrated 
luminosities 
of 10\,pb$^{-1}$, 10s of millions of events are expected.  It should be noted that the lowest value of $Q^2$ measured at HERA 
was $Q^2 = 0.045$\,GeV$^2$, which at VHEeP corresponds to a minimum $x$ value of $5 \times 10^{-10}$.  At this $Q^2$, a 
significantly larger number of events is expected.  Hence high precision measurements with negligible statistical uncertainties 
will be possible at VHEeP.  Also shown in Fig.~\ref{fig:vheep_kinematics} are isolines for the angles of the scattered electron and 
final-state hadronic system.  This highlights the need for instrumentation close to the beam-pipe in the direction of the electron 
beam in order to be able to measure the scattered electron at low $x$.  It also highlights the need for hermetic instrumentation 
to measure the hadronic final state where events at low $x$ have a hadronic system at low angles in the direction of the 
electron beam.  Conversely, events at high $x$ have a hadronic system at low angles in the direction of the proton beam.  
Clearly the detector design for VHEeP will have a number of challenges and will need to be different from conventional 
collider experiments such as at the LHC.

\begin{figure}[!h]
\centering\includegraphics[width=12cm]{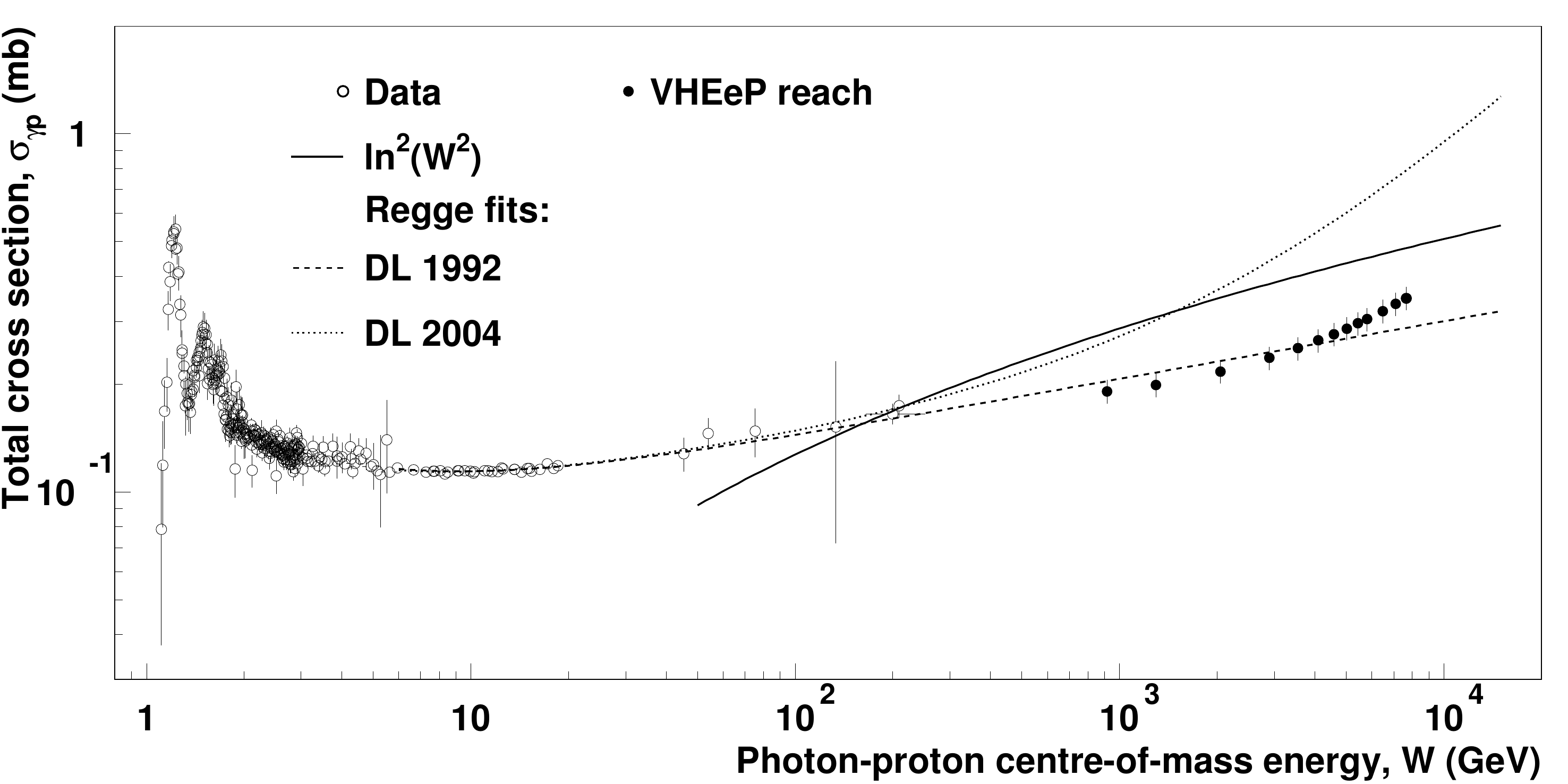}
\caption{Total $\gamma p$ cross section versus photon--proton centre-of-mass energy, $W$, shown for data compared to 
various models~\cite{pl:b296:227,Donnachie:2004pi,pr:123:1053}.  The data is taken from the PDG~\cite{pdg}, with 
references to the original papers given therein. The highest VHEeP data point is shown at $y=0.7$ (where $y = W^2/s$); the 
cross section at this point is assumed to be double the ZEUS value.  The other VHEeP points assume different values of 
$y$  down to 0.01 and are plotted on a straight line (linear in $W$) between the ZEUS and highest VHEeP point.  All VHEeP 
points have the same uncertainty as the ZEUS point: a systematic uncertainty of 7.5\% and a negligible statistical uncertainty.  
The ZEUS measurement is at $\sqrt{s} = 209$\,GeV and used a luminosity of 49\,nb$^{-1}$. This result has been updated 
from the original paper with the addition of newly-calculated points for VHEeP.}
\label{fig_sigtot}
\end{figure}

The physics potential of VHEeP was discussed in the original publication~\cite{epj:c76:463}.  An example and recently 
updated result is shown in Fig.~\ref{fig_sigtot}, 
in which the total $\gamma p$ cross section is shown versus the photon--proton centre-of-mass energy, $W$.  This is 
a measurement which relies on only a modest luminosity and will be dominated by systematic uncertainties.  As can 
be seen from the expected VHEeP data, the measurement is extended to energies well beyond the current data, into 
a region where the dependency of the cross section is not known.  Some models are also shown and they clearly 
differ from each other at the high energies achievable at VHEeP.  These data could also be useful in understanding 
more about cosmic-ray physics as such collisions correspond to a 20\,PeV photon on a fixed target.

The energy dependence of scattering cross sections for virtual photons on protons is also of fundamental interest, and its study 
at different virtuality is expected to bring insight into the processes leading to the observed universal behaviour of cross sections 
at high energies.  In deep inelastic scattering of electrons on protons at HERA, the strong increase in the proton structure function $F_2$ with 
decreasing $x$ for fixed, large, $Q^2$ is usually interpreted as an increasing density of partons in the proton, providing more 
scattering targets for the electron.  This interpretation relies on choosing a particular reference frame to view the scattering -- 
the Bjorken frame.  In the frame where the proton is at rest, it is the state of the photon or weak boson that differs with varying 
kinematic parameters.  For the bulk of the electron--proton interactions, the scattering process involves a photon, and we can 
speak of different states of the photon scattering on a fixed proton target.  What is seen is that the photon--proton cross section 
rises quickly with $W$ for fixed $Q^2$~\cite{ref:sigmagp}.  In the proton rest 
frame, we interpret this as follows: as the energy of the photon increases, time dilation allows shorter lived fluctuations of the 
photon to become active in the scattering process, thereby increasing the scattering cross section.  

\begin{figure}[!h]
\centering\includegraphics[width=12cm]{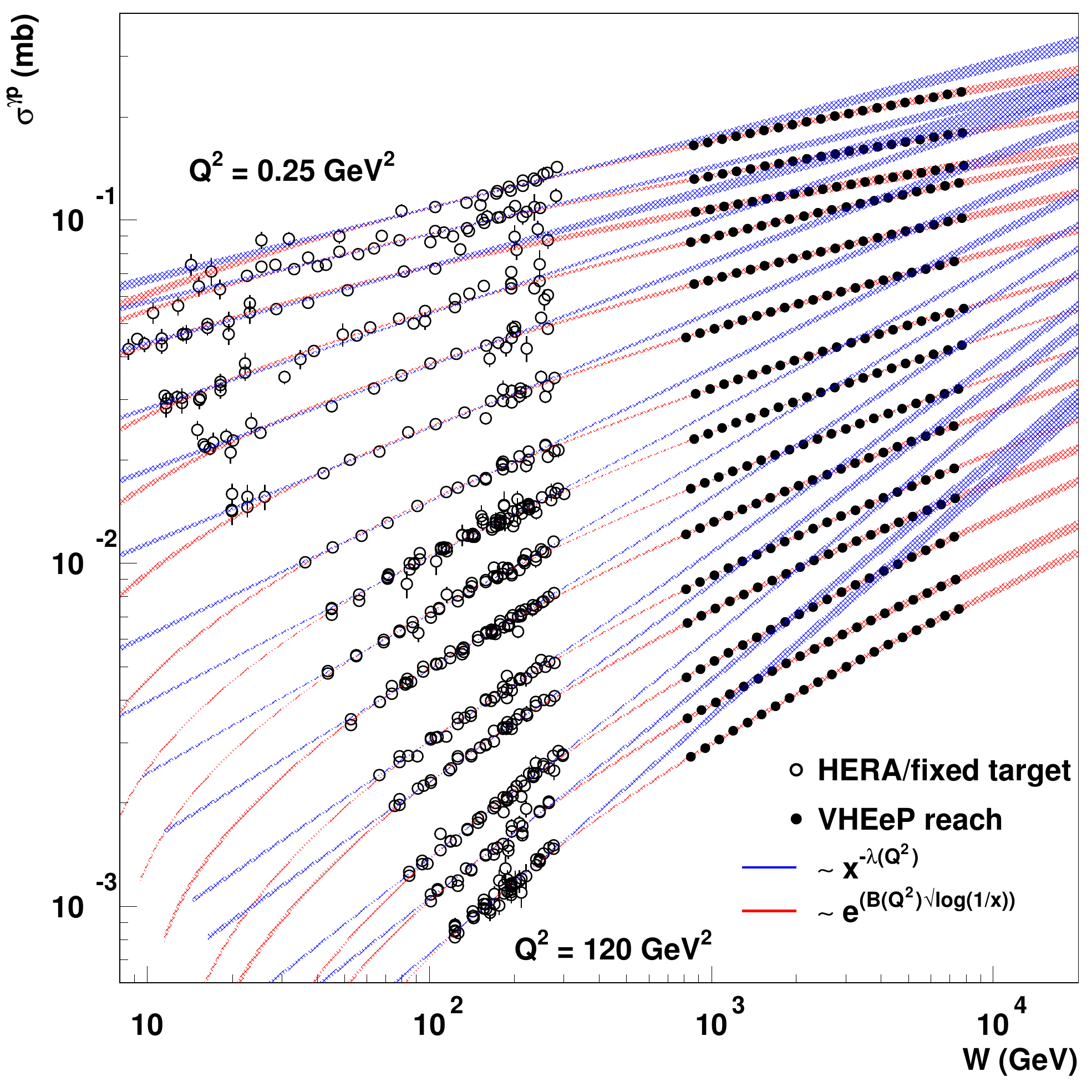}
\caption{Measurements (open points) of $\sigma^{\gamma p}$ versus $W$ for $0.25 < Q^2 < 120$\,GeV$^2$ from HERA 
and fixed-target experiments.  The blue and red lines show fits to the data, performed separately for each $Q^2$ value, of 
the forms given in the key.  The reach of VHEeP is shown as projected data points (closed points).  The points are placed 
on the red curve.  The uncertainties are assumed to be of order 1\%, given the increased cross section expected and 
similar systematics to those at HERA and are not visible as error bars on this plot.}
\label{fig_sig-gammap}
\end{figure}

Figure~\ref{fig_sig-gammap} shows the results of extrapolation of fits to the energy dependence of the photon--proton cross section 
for different photon virtualities as given in the caption~\cite{ref:caldwell}, for two different assumptions on the energy behaviour.  It is 
found that the simple behaviour 
cannot continue to ever smaller values of $x$ as this would result in large-$Q^2$ cross sections becoming larger than 
small-$Q^2$ cross sections.  A change of the energy dependence is therefore expected to become visible in the VHEeP 
kinematic range.  This should yield exciting and unique information on the fundamental underlying physics at the heart 
of the high energy dependence of hadronic cross sections.

At the very highest $Q^2$ values, searches for high energy phenomenon beyond the Standard Model will be possible.  Higher luminosities 
will allow a comprehensive search to complement those at the LHC; however, even with modest luminosities, some specific processes can 
be investigated with higher sensitivity at VHEeP than at the LHC.  As an example, the production of leptoquarks, which would be produced 
on mass shell, is possible up to the kinematic limit of the centre-of-mass energy, i.e.\ 9\,TeV.  Other examples of physics that could be 
investigated at VHEeP were presented and discussed at a workshop\cite{vheep-ws}.  It was discussed how high-energy $ep$ 
collisions are sensitive to new descriptions and general theories of particle interactions~\cite{dvali} as well as having connections with 
black holes and gravity~\cite{erdmenger}.  

The work presented here has focused on the use of protons as wakefield drivers and as the collision particles, i.e.\ $ep$ 
physics, however, the studies will be extended to consider ions as both wakefield drivers and collision particles.  Ions, in 
particular those which are partially stripped (this technique is also discussed as part of the Physics Beyond Colliders 
study)~\cite{krasny}, could be significantly cooled such that the bunches are more intense and so will be more effective 
wakefield drivers and lead to much higher luminosities for $eA$ collisions.

\section{Summary}

The AWAKE scheme is a promising method to accelerate electrons to high energy over short distances using proton-driven plasma 
wakefield acceleration.  The AWAKE collaboration has already demonstrated strong fields and electron acceleration and is now 
embarking on a new programme to demonstrate this as a useable technology for particle physics experiments.  This document 
outlines some of these applications and shows that the AWAKE technology could lead to unique facilities and experiments 
that would otherwise not be possible.  In particular, experiments are proposed to search for dark photons, measure strong field 
QED and investigate new physics in electron--proton collisions.  The community is also invited to consider applications for 
electron beams up to the TeV scale.

\section*{Acknowledgements}

This work was supported in parts by: the Leverhulme Trust Research Project Grant RPG-2017-143; and the Swiss National Science 
Foundation (SNSF) Grant No.\ 169133 and ETH Z\"{u}rich (Switzerland).  M.Wing acknowledges the support of DESY, Hamburg.

%% file: cernrep.bbl
\begin{thebibliography}{99}
\setlength{\itemsep}{-0.mm} 

\bibitem{AWAKE-sub}
A. Caldwell et al., AWAKE -- on the path to particle physics applications, Input to the European Particle Physics Strategy Update 
2018--2020 (2018).

\bibitem{pp:18:103101}
A. Caldwell and K. Lotov, 
\textit{Phys. Plasmas} {\bf 18} (2011) 103101.

\bibitem{wing:royalsoc}
M. Wing, 
arXiv:1810.12254.

\bibitem{AWAKE-technical}
E. Gschwendtner et al., AWAKE++: the AWAKE acceleration scheme for new particle physics experiments at CERN, Input to the 
European Particle Physics Strategy Update CERN-PBC-REPORT-2018-005 (2018).

\bibitem{awake1}
AWAKE Coll., E. Adli et al., 
arXiv:1809.04478.

\bibitem{awake2}
AWAKE Coll., M. Turner et al., 
arXiv:1809.01191.

\bibitem{awake3}
AWAKE Coll., E. Adli et al., 
\textit{Nature} \textbf{561} (2018) 363.

\bibitem{muon-collider}
M. Antonelli et al., 
\textit{Nucl. Instrum. Meth.} \textbf{807} (2016) 101.

\bibitem{fcc}
A. Ball et al., 
CERN EDMS-1342402 (2014).

\bibitem{JETP:56:502}
L.B. Okun, 
\textit{Sov. Phys. JETP} \textbf{56} (1982) 502 [\textit{Zh. Eksp. Teor. Fiz.} \textbf{83}, 892].

\bibitem{pl:b136:279}
P. Galison and A. Manohar, 
\textit{Phys. Lett.} \textbf{B~136} (1984) 279.

\bibitem{pl:b166:196}
B. Holdom, 
\textit{Phys. Lett.} \textbf{B~166} (1986) 196.

\bibitem{pl:b513:119}
S.N.~Gninenko and N.V. Krasnikov, 
\textit{Phys. Lett.} \textbf{B~513} (2001) 119.

\bibitem{pr:d79:015014}
N. Arkani-Hamed et al., 
\textit{Phys. Rev.} \textbf{D~79} (2009) 015014.

\bibitem{arnps:60:405}
J. Jaeckel and A. Ringwald, 
\textit{Ann. Rev. Nucl. Part. Sci.} \textbf{60} (2010) 405.

\bibitem{us-cosmic-visions}
M. Battaglieri et al., 
arXiv:1707.04591.

\bibitem{pr:d89:075008}
S. Gninenko, 
\textit{Phys. Rev.} \textbf{D~89} (2014) 075008.

\bibitem{arxiv:1312.3309}
S. Andreas et al., 
arXiv:1312.3309.

\bibitem{prl:118:011802}
NA64 Coll., D. Banerjee et al., 
\textit{Phys. Rev. Lett.} \textbf{118} (2017) 011802.

\bibitem{geant}
S. Agostinelli et al., 
\textbf{A~506} (2003) 250.

\bibitem{hartin}
A. Hartin, 
\textit{Int. J. Mod. Phys.} \textbf{A~33} (2018) 1830011.

\bibitem{pr:d60:092004}
E144 Coll., C. Bamber et al., 
\textit{Phys. Rev.} \textbf{D~60} (1999) 092004.

\bibitem{sqed-ws}
See talks at workshop on "Probing strong-field QED in electron--photon interactions" (2018) DESY, 
{\tt https://indico.desy.de/indico/event/19493/}

\bibitem{arXiv:1807.10670}
A. Hartin, A. Ringwald and N. Tapia, 
arXiv:1807.10670.

\bibitem{dis:book}
See for example: R. Devenish and A. Cooper-Sarkar, \textit{Deep Inelastic Scattering}.  Oxford University Press, Oxford, UK (2004).

\bibitem{hera}
HERA -- A Proposal for a Large Electron Proton Colliding Beam Facility at DESY. 1981. DESY-HERA-81-10.

\bibitem{mpl:a24:1087}
See for example: S.D.~Bass, 
\textit{Mod. Phys. Lett.} \textbf{A~24} (2009) 1087.

\bibitem{guoxing}
G. Xia et al., 
\textit{Nucl. Instrum. Meth.} \textbf{A~740} (2014) 173.

\bibitem{epj:c76:463} 
A. Caldwell and M. Wing, 
\textit{Eur. Phys. J.} \textbf{C~76} (2016) 463.

\bibitem{pl:b296:227}
A. Donnachie and P.V. Landshoff, 
\textit{Phys. Lett.} \textbf{B~296} (1992) 227.

\bibitem{Donnachie:2004pi}
A. Donnachie and P.V. Landshoff, 
\textit{Phys. Lett.} \textbf{B~595} (2004) 393. 

\bibitem{pr:123:1053}
M. Froissart, 
\textit{Phys. Rev.} \textbf{123} (1961) 1053.

\bibitem{pdg}
Particle Data Group, Yao W-M, et al., 
\textit{J. Phys.} \textbf{G~33} (2006) 1; data from 
{\tt http://pdg.lbl.gov/2006/hadronic-xsections/hadron.html}

\bibitem{ref:sigmagp}
ZEUS Collaboration, J. Breitweg et al., \textit{Phys. Lett.} {\bf B~407} (1997) 432.

\bibitem{ref:caldwell}
A. Caldwell, 
\textit{New J. Phys.} {\bf 18} (2016) 073019.

\bibitem{vheep-ws}
See talks at workshop on "Prospects for a very high energy $eP$ and $eA$ collider and Leo Stodolsky symposium", 
Max Planck Institute for Physics, Munich (2017) {\tt https://indico.mpp.mpg.de/event/5222/overview}

\bibitem{dvali}
G. Dvali, High energy cross sections and classicalization. Talk presented at~\cite{vheep-ws} (2017).

\bibitem{erdmenger}
J. Erdmenger, Applications of AdS/CFT to very low-$x$ physics. Talk presented at~\cite{vheep-ws} (2017).

\bibitem{krasny}
M.W. Krasny, 
arXiv:1511.07794.

\end{thebibliography}
